\pdfoutput=1
\documentclass[12pt]{iopart}

\usepackage{amsfonts}
\usepackage{amssymb}
\usepackage{graphicx}
\usepackage{epstopdf}
\usepackage{verbatim}
\usepackage{color}
\usepackage{multirow}
\usepackage{mathcomp} 
\usepackage{array}
\usepackage{bm}
\usepackage{wasysym}
\usepackage{setstack}
\usepackage[utf8]{inputenc}
\usepackage{csquotes}
\usepackage{harvard}
\usepackage[thickspace,amssymb]{SIunits}

\newcommand{\he}{$^{\mathrm{4}}$He}
\newcommand{\car}{$^{\mathrm{12}}$C}
\newcommand{\oxi}{$^{\mathrm{16}}$O}

\begin{document}

\title[]{Secondary radiation measurements for particle therapy applications: prompt photons produced by $^{4}$He, $^{12}$C and $^{16}$O ion beams in a PMMA target}

\author{I.~Mattei$^{a}$, F.~Collamati$^{b}$,
  E.~De Lucia$^{b}$, R.~Faccini$^{c,d}$, P.~M.~Frallicciardi$^{e}$,
  C.~Mancini-Terracciano$^{c,d}$,
  M.~Marafini$^{d,f}$, S.~Muraro$^{a}$, R.~Paramatti$^{d}$,
  V.~Patera$^{d,f,g}$, L.~Piersanti$^{d,g}$, D.~Pinci$^{d}$,
  A.~Rucinski$^{d,g}$, A.~Russomando$^{c,d,h}$, A.~Sarti$^{b,f,g}$,
  A.~Sciubba$^{d,f,g}$, E.~Solfaroli~Camillocci$^{c,d}$, M.~Toppi$^b$,
  G.~Traini$^{c,d}$, C.~Voena$^{d}$, G.~Battistoni$^{a}$}

\address{$^a$ INFN - Sezione di Milano, Italy}
\address{$^b$ Laboratori Nazionali di Frascati dell'INFN, Frascati, Italy} 
\address{$^c$ Dipartimento di Fisica, Sapienza Universit\`a di Roma, Roma, Italy}
\address{$^d$ INFN - Sezione di Roma, Italy}
\address{$^e$ Istituto di ricerche cliniche Ecomedica, Empoli, Italy}
\address{$^f$ Museo Storico della Fisica e Centro Studi e Ricerche ``E.~Fermi'', Roma, Italy}
\address{$^g$ Dipartimento di Scienze di Base e Applicate per Ingegneria, Sapienza Universit\`a di Roma,  Roma, Italy}
\address{$^h$ Center for Life Nano Science@Sapienza, Istituto Italiano di Tecnologia, Roma, Italy}

\begin{abstract}


Charged particle beams are used in Particle Therapy (PT) to treat oncological patients due to 
their selective dose deposition in tissues with respect to photons and electrons used in conventional radiotherapy. Heavy (Z$>$1) PT beams can additionally exploit their high biological effect in killing cancer cells. Nowadays, protons and carbon ions are used in PT clinical routine. 
Recently, the interest on the potential application of helium and oxygen beams is growing: with respect  to protons such beams are characterized by their reduced multiple scattering inside the body and increased linear energy transfer, relative biological effectiveness and oxygen enhancement ratio.\\ 

The precision of PT demands for online dose monitoring techniques, crucial to improve the quality assurance of treatments: possible patient mis-positionings and tissues biological changes with respect to the CT scan could negatively affect the therapy outcome.
The beam range confined in the irradiated target can be monitored thanks to the neutral or charged secondary radiation emitted by the interactions of hadron beams with matter. Among these secondary products, prompt photons are produced by nuclear de-excitation processes and, at present, different dose monitoring and beam range verification techniques based on the prompt-$\gamma$ detection have been proposed. It is hence of importance to perform the $\gamma$ yield measurement in therapeutical-like conditions.

In this paper we report the yields of prompt photons produced by the interaction of helium, 
carbon and oxygen ion beams with a poly-methyl methacrylate (PMMA) beam stopping target. 
The measurements were performed at the Heidelberg Ion-Beam Therapy center (HIT) with 
beams of different energies. A LYSO scintillator, placed at $60\degree$ 
and $90\degree$ with respect to the beam direction, has been used as photon detector. 
The obtained $\gamma$ yields for carbon ion beams are compared with results from literature, while no other results from helium and oxygen beams have been published yet. A discussion on the expected resolution of a slit camera detector is presented, demonstrating the feasibility of a prompt-$\gamma$ based monitoring technique for PT treatments using helium, carbon and oxygen ion beams.\\

\end{abstract}



\section*{Introduction}

Particle Therapy (PT) exploits the characteristic energy release in matter of charged particles, mainly protons and carbon ions, in the irradiation of tumor volumes trying to spare as much as possible the surrounding healthy tissues. In comparison to the most advanced photon radiotherapy technique, protons and carbon ions have the advantageous feature of a high Linear Energy Transfer (LET) and increased Relative Biological Effectiveness (RBE), which make those particles particularly favourable in treating radio-resistant tumors~\cite{Loeffer13}. Beside protons and carbon ions, helium and oxygen ion beams are currently investigated as PT candidates~\cite{TommasinoHeO}: helium ions suffer less lateral multiple scattering with a consequent reduced beam broadening with respect to protons, providing a good solution in the irradiation of radio-resistant tumors when low beam fragmentation is required
~\cite{Fuchs}. Oxygen ions could be of advantage when increasing efficiency in treating radio-resistant tumors is needed due to their higher LET, RBE and Oxygen Enhancement Ratio (OER) with respect to protons, carbons and heliums~\cite{Kurz}.

The high selectivity of PT makes this technique particularly sensitive to possible patient mis-positionings and anatomical variations, asking for the development of an online beam range monitor. This should be able to provide a feedback on the dose deposition spatial distribution during the treatment in order to improve its
quality and efficacy: the lack of a precise online monitoring is one of 
the key issues to be addressed to support the diffusion of PT
therapies in clinical centers. 
The beam range monitoring takes advantage of the secondary radiation 
produced by the interactions of the beam with the target
nuclei along the path inside the target volume: the beam is
stopped inside the patient and the secondaries escaping from the body
can be detected, their emission profile reconstructed and used to monitor the beam 
dose deposition and range. 
So far, the most established PT beam monitoring technique is based on the detection of back-to-back photons produced by the annihilation positrons coming from $\beta^+$ emitters (PET photons) using the Positron Emission Tomography (PET) technique. Nevertheless, the signal level is lower in comparison with PET signals known from clinical diagnostics, and at present such technique is used off-line, after the patient irradiation, aiming for further investigation methods~\cite{Paro13}.\\
A technology capable of on-line PET detection~\cite{Pawelke1997,Parodi2002,Priegnitz2008,Fiedler2008} is 
currently under development~\cite{Attanasi2009,Vecchio2009,Inside,Inside2}.
The characterization of the secondary radiation
produced by p and \car~beams of therapeutical energy has been the subject 
of an intensive experimental campaign in the recent past: the
production of PET photons~\cite{Enghardt2004,Agodi2012Beta}, 
light charged  fragments~\cite{Agodi2012CP,Henriquet2012,Gwosch2013,Piersanti2014} and
prompt photons, main object of this contribution, has been studied in different experimental 
conditions~\cite{aafke2015}.
In particular, the prompt-$\gamma$ production has been recently investigated by several
experiments performed using p and \car~beams of several energies: 
$73\ \mega\electronvolt/\text{u}$~\cite{Testa2008,Testa2009}, 
$80\ \mega\electronvolt/\text{u}$~\cite{Agodi2012PP,AgodiERRATA}, 
$220\ \mega\electronvolt/\text{u}$~\cite{GsiLYSO220},
$95\ \mega\electronvolt/\text{u}$, 
$300\ \mega\electronvolt/\text{u}$ and 
$310\ \mega\electronvolt/\text{u}$~\cite{Pinto2015,Testa2010}. The prompt-$\gamma$ yield 
estimation is of particular importance since it strongly affects the achievable resolution on the evaluation 
of the hadron beam range. The main interests
in using this kind of secondary radiation for monitoring purposes are related to 
its large abundance and the high 
precision achievable on the emission point reconstruction 
in absence of multiple scattering interactions. \\ 

In this paper we present the results of prompt photons production 
measurements performed using \he, \car~and \oxi~beams,
available at the Heidelberg Ion-beam Therapy center (HIT, Heidelberg, 
Germany), interacting with a beam stopping PMMA target.  

The experimental setup used for the data acquisition is described in 
detail in \S~\ref{exp-setup}, while the results obtained with the LYSO detector at
$60\degree$ and $90\degree$ with respect to the beam direction are
discussed in \S~\ref{Yields} and compared with the results available from literature. 
The discussion of the measured fluxes in the context of monitoring applications 
can be found in \S~\ref{monitor}, taking as a reference detector for
prompt gamma monitoring the IBA knife edge slit camera. 
Particular care has been made to quantify the impact of the 
collectable statistics of prompt photons emitted in 
Particle Therapy with heavy ions clinical like scenarios on the 
precision achievable on the dose deposition monitoring.

The experimental apparatus did not allow a measurement of the 
prompt photons production position and hence a disentangling 
of the photons production induced by the primary incoming radiation
and of the secondary photons produced after the Bragg Peak region
was not possible. The implications for the prompt photons monitoring
are detailed in \S~\ref{monitor}.


\section{Experimental setup}
\label{exp-setup}

The experiment was performed at the HIT center where
different ion species of several energies were used to irradiate 
a poly-methyl methacrylate (PMMA) target to study the prompt-$\gamma$ radiation emitted from the interactions of the beam 
projectiles with the target nuclei.\\

\begin{figure}[t]
\centering
\includegraphics[width = 0.6\textwidth]{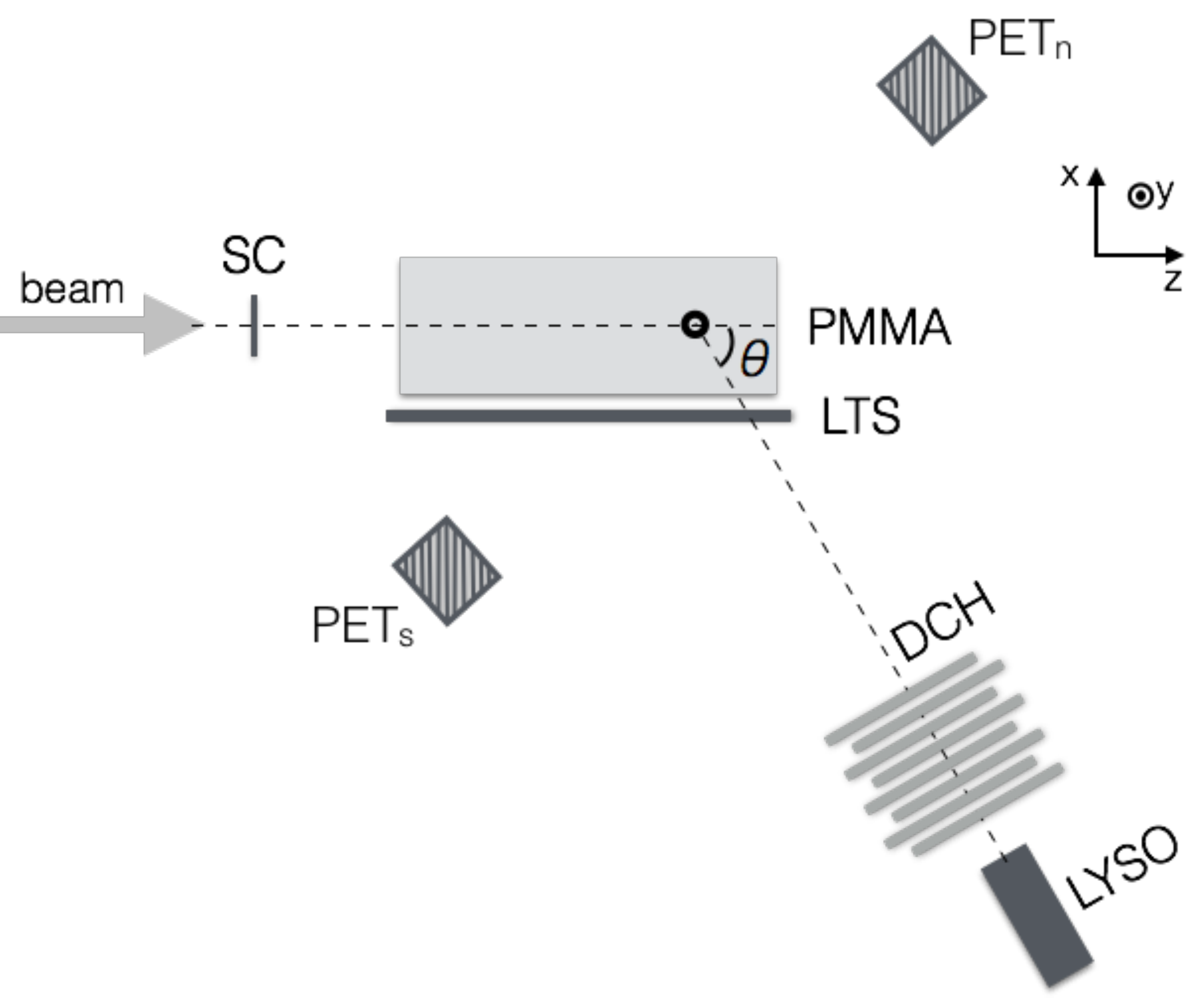}
\caption{Experimental setup scheme (not to scale). The LYSO and 
DCH detectors were mounted on a movable arm that allowed to
detect the secondary radiation in two different angular
configurations:  $\theta=60\degree$ or $90\degree$, where $\theta$ is
the angle with respect to the beam direction in the
horizontal plane. $\gamma$-PET detectors (PET$_s$ and PET$_n$)
are also shown.}
\label{fig:expset}
\end{figure}

Figure~\ref{fig:expset} shows the experimental setup, which is part of a more extended geometry meant to measure not only the prompt photon radiation, but also secondary charged particles, PET photons and forward emitted heavy fragments. 

The origin 
of the reference frame is marked by
the black spot inside the PMMA target.
The beam, coming from the left along the z axis, is illustrated with 
an arrow. To detect the incoming primary particles, a plastic 
scintillator (EJ$200$) $2\ \milli\meter$ thick (Start Counter, SC) 
was placed $\sim 30\ \centi\meter$ upstream the PMMA. 
The SC was read by two photomultiplier 
tubes (PMTs) - Hamamatsu H$6524$ - each one giving a recorded signal (SC1, SC2). 
The SC, used as a trigger detector, provided the number of impinging ions and the 
reference time for the Time of Flight (ToF) measurements.

The PMMA target dimensions are $(5.00 \times 5.00 
\times \text{t}_{\rm PMMA})\ \centi\meter^3$, where t$_{\rm PMMA}$ 
is the PMMA thickness that depends on the penetration depth of the primary beam 
and experimental setup configuration used during the data acquisition
(see Table~\ref{tab:datatake}). The uncertainty on t$_{\rm PMMA}$ is $0.01\ \centi\meter$. For the \he~and \oxi~ions data collection, in order to keep the Bragg Peak (BP) position fixed for all the beam energies,
the size of the PMMA was changed according to the energy and range (BP depth) of each beam. For the carbon ion data acquisition the PMMA thickness 
was fixed at $10.00\ \centi\meter$ regardless of the incoming
beam energy, still ensuring that the beam stops within the PMMA.

A Long Thin Scintillator (LTS) $(0.2 \times 5.0 \times 17.0)\ \centi\meter^3$, 
read by a H$10580$ PMT, was placed on the PMMA lateral face in order to 
identify and study secondary charged particles. 
A $21\ \centi\meter$ long Drift CHamber (DCH) was mechanically aligned 
with the reference frame origin and was placed at a distance of 
$\sim 60\ \centi\meter$ from the PMMA center. 
The DCH, described in details
in~\cite{AbouHaidar,Agodi2012CP,Piersanti2014}, 
was used to identify and reject the 
background from low energy charged particles (see \S~\ref{Yields}).

Prompt photons were detected by a $2\times 2$ matrix of scintillating lutetium yttrium orthosilicate (LYSO) crystals, $(1.5\times1.5\times 12.0)\ \centi\meter^3$ each, described 
in detail in~\cite{Agodi2012PP}. The detector was placed $\sim 2\
\centi\meter$ behind the DCH exit face and was mechanically 
aligned with the reference frame origin.
In order to discriminate the LYSO signal from the LYSO intrinsic 
background, 
a PMT threshold has been set
corresponding to a calibrated LYSO deposited energy of 
$\sim 1\ \mega\electronvolt$. 

 The energy calibration of the LYSO scintillator was performed 
with the final data acquisition setup at the HIT center using a 
radioactive source of $^{60}\text{Co}$. The data collected
have been used to provide the reference point needed
for the implementation of a former calibration of the 
very same crystal that was performed in the range of interest for prompt
photons emission studies, with a different experimental setup, 
and is documented in detail in a separate manuscript~\cite{Bellini2014}.
A linear calibration curve, up to $\sim10\ \mega\electronvolt$,
is assumed following the results obtained
in the extended calibration energy range study presented
in~\cite{Bellini2014}. 
The ToF slewing effect induced by the front-end electronics fixed voltage
threshold was taken into account following 
the procedure described in~\cite{Agodi2012PP,AgodiERRATA}.

The DCH and LYSO detectors were attached to the same movable aluminum arm. This arm 
could be placed at different angles with respect to the beam  direction ($\theta$),  \textit{i.e.} $60\degree$ and
$90\degree$, in order to measure the angular dependence of the prompt-$\gamma$ emission.
The total distance of the LYSO
from the PMMA center was $(82.0 \pm 0.1)\ \centi\meter$ in the $90\degree$ setup and $(86.5 \pm 0.1)\ \centi\meter$ in the $60\degree$ setup.

Fig.~\ref{fig:expset} shows also the positioning of two LYSO
pixellated $\gamma$-PET detectors
(PET$_s$ and PET$_n$) used to monitor the production of $beta^+$ 
emitters\footnote{The obtained results, as well as the experimental methods, 
will be the subject of a dedicated manuscript,
currently in preparation, and 
will be hence documented in detail elsewhere.}
during the PMMA irradiation. The PET trigger
line required the coincidence in time of the signals
of the LYSO detectors and was independent from the SC signals
in order to allow the measurements of the $\beta^+$ induced
activity also when the beam was not impinging on the target.

The $\gamma$-PET detectors were included in the simulation of
the experimental apparatus and used to provide an independent check on the
number of incoming ions, as described in \S~\ref{res:sys}.

\begin{table}[ht!]
\centering
{
\renewcommand\arraystretch{1.1}
\centering
\begin{tabular}{c c c c c c}
\textbf{Ion} &\textbf{Energy} & \textbf{B$_{\rm \textbf{FWHM}}$ } & \textbf{Range} & \textbf{t$_{\rm \textbf{PMMA}}$}
& $\bm{ \theta}$ \\
&  (MeV/u) & (mm) & (cm) & (cm) &  \\
\hline  
\multirow{4}* {$^{12}{\rm C}$} & $120.45$ & 7.9 & 2.88 & \multirow{4}* {10.00} & \multirow{4}*{$90\degree$}\\ 
 & $159.99$ & 6.2 & 4.83 & \\
& $180.89$ & 5.5 & 6.03 & \\
& $219.79$ & 4.7 & 8.33 & \\
\hline
\multirow{3}* {$^{4}{\rm He}$} & 102.34 & 9.3 & 6.68 & 7.65 & $60\degree$ \\
\cline{6-6}
 & $124.78$ & 7.8 & $9.68$ & $10.00$ & \multirow{2}*{$90\degree$\ -\ $60\degree$} \\ 
& $144.63$ & 6.9  & 12.63 & 12.65 & \\ 
\hline
\multirow{3}*{$^{16}{\rm O}$} & $209.63$ & 4.6 & 5.78 & 7.65 & \multirow{3}*{$90\degree$\ -\ $60\degree$}\\
& $259.55$ & 3.9 & 8.38  & 10.00 & \\
 & $300.13$ & 3.6 & 10.68 & 12.65 & \\
\hline
\end{tabular}
}
\caption{The beam energy, beam spot size (FWHM), beam range (BP depth) and PMMA thickness used during
the data acquisition are listed as a function of the beam ion species 
and angular configuration ($\theta = $ angle at which the DCH and LYSO
detector were placed with respect to the beam direction). }
\label{tab:datatake}
\end{table}

Table~\ref{tab:datatake} summarizes the measured setup configurations
relative to the collected data sample. For each ion species, the beam energy (\textit{Energy}) and spot size (\textit{B$_{FWHM}$}) as the Full Width at Half Maximum (FWHM) from the HIT libraries are reported (later in the text, the beam energies will be approximated values). The relative error on the HIT beam energies is of $1.5\cdot 10^{-3}$~\cite{ParodiHIT}. The beam range in PMMA, defined as the BP position depth inside the target (\textit{Range}), was computed using a 
FLUKA Monte Carlo (MC) simulation~\cite{Ferrari2005,Boehlen2014,Battistoni2015}. The error on the reported \textit{Range} values is $0.05\ \centi\meter$, determined from the simulated Bragg curves of each beam impinging on a PMMA target (density: $1.19\ \gram~\centi\meter^{-3}$; ionisation potential: $74\ \electronvolt$).  
For each data taking, the PMMA thickness (\textit{t$_{PMMA}$}) and the DCH - LYSO angular configuration ($\theta$) are also listed. 

The Data AcQuisition (DAQ) trigger was provided by the time coincidence of the logic OR of the SC signals 
(SC1 or SC2) with the LYSO signal. This choice was driven by the need to optimize the
data sample collection efficiency and to allow for a measurement of the SC 
detector efficiency. It has to be noted that the data analysis performed for
the prompt photons study requires the AND of the SC1, SC2 signals in all the steps, 
to minimize the contribution of random electronic noise to the measurement
of the number of incoming ions.

All electronic signals were read-out by a VME system (CAEN V2718 VME-PCI bridge) 
interfaced with a PC for the DAQ. 
The time and charge information for the signals of all the detectors
were acquired using a $19$-bit TDC Multi-hit 
(CAEN V1190B, time resolution 
of $\sim100\ \pico\second$), and a $12$-bit QDC 
(CAEN V792N,  resolution of
$\sim0.1\ \pico\coulomb$). 
The digitization of the time information was accomplished by means of a 
discrimination system in which the digital output signals had a $100\ \nano\second$ 
width. The impact of this choice in the counting of events with multiple
ions impinging on the SC is discussed in detail below (\S~\ref{beam}).

The number of impinging ions was 
counted by means of a VME scaler (CAEN V560 N), using the logic AND of the SC signals 
(SC1 and SC2, defined as SCAND). The VME scaler is capable of sustaining an incoming signal rate up to $100\ \mega\hertz$.






\subsection{Beam description}
\label{beam}

The ion beams provided by the HIT facility are accelerated using a
synchrotron. The incoming beam rate,
kept under the $10\ \mega\hertz$ limit set by the SC signal
discrimination time used for the ToF measurements, ranged from $\sim 300\
\kilo\hertz$ up to $\simeq 3\ \mega\hertz$, depending on 
the ion beam species. Such rate has been heavily reduced with respect to the therapeutical rates in order to 
allow the experimental data taking. 

\begin{figure}[h]
\centering
\begin{minipage}{0.33\linewidth}
\centering
\includegraphics[width = 1  \textwidth]{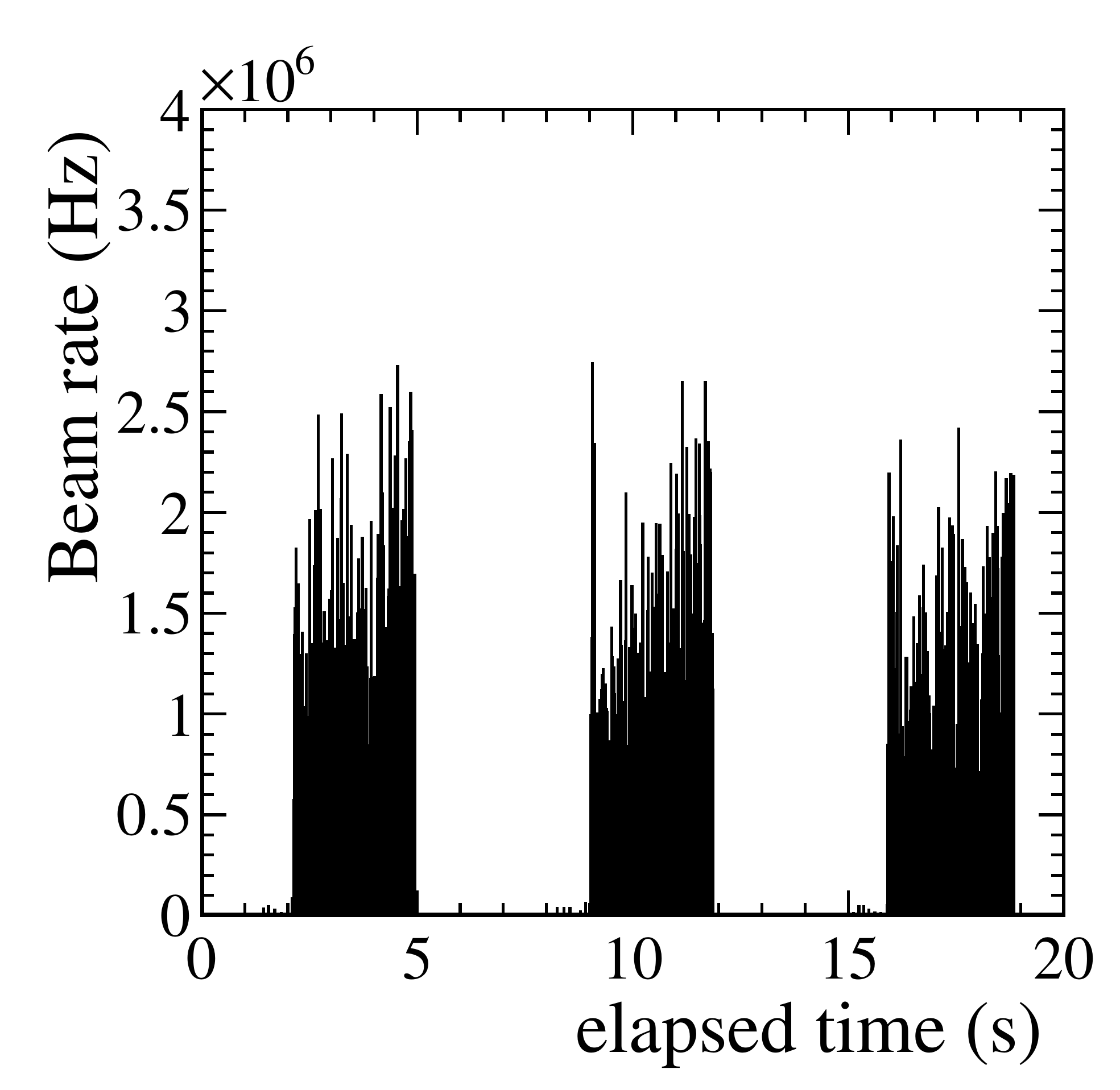}
\end{minipage}\hfill
\begin{minipage}{0.33\linewidth}
\centering
\includegraphics[width = 1  \textwidth]{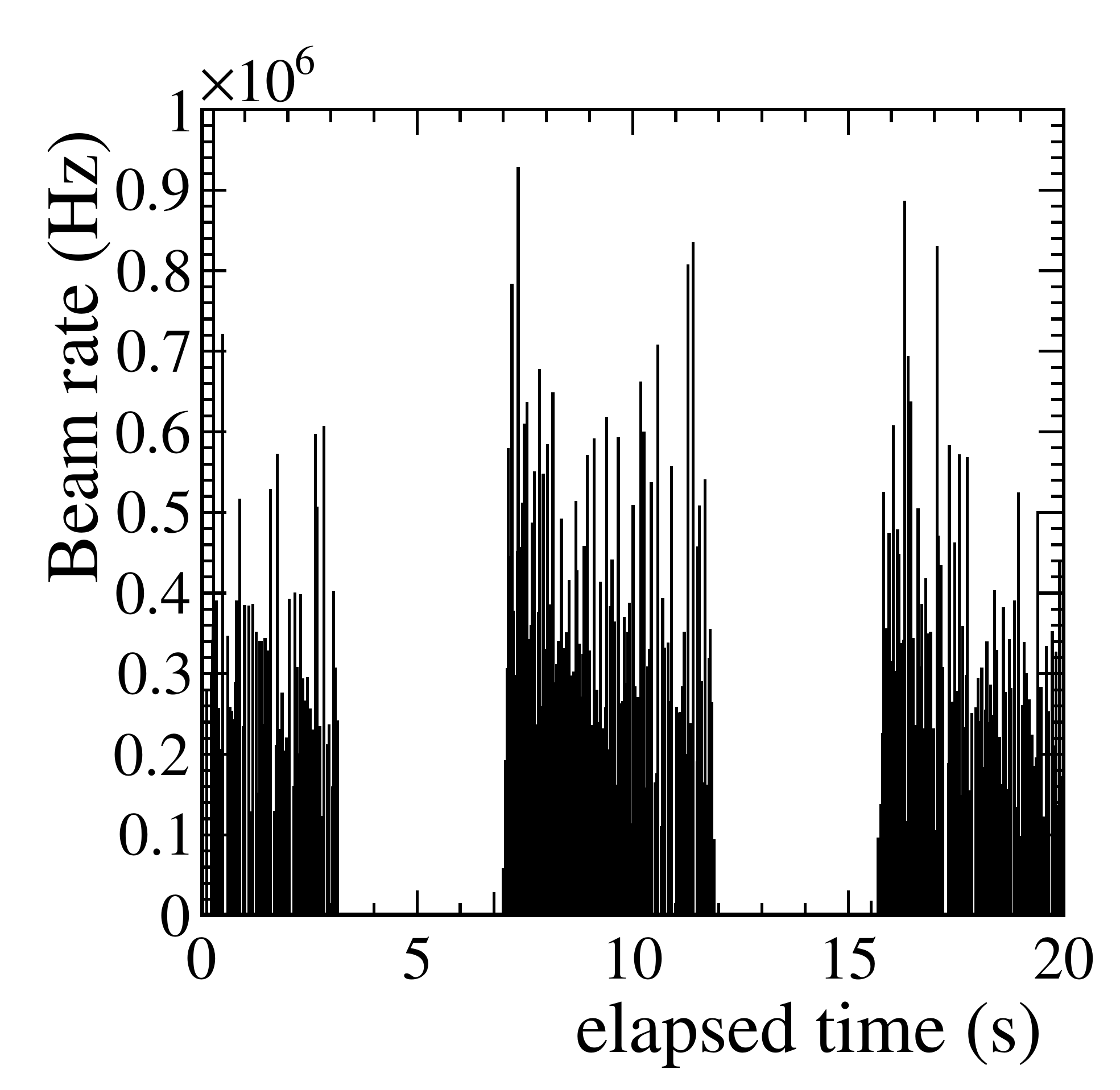}
\end{minipage}
\begin{minipage}{0.33\linewidth}
\centering
\includegraphics[width = 1  \textwidth]{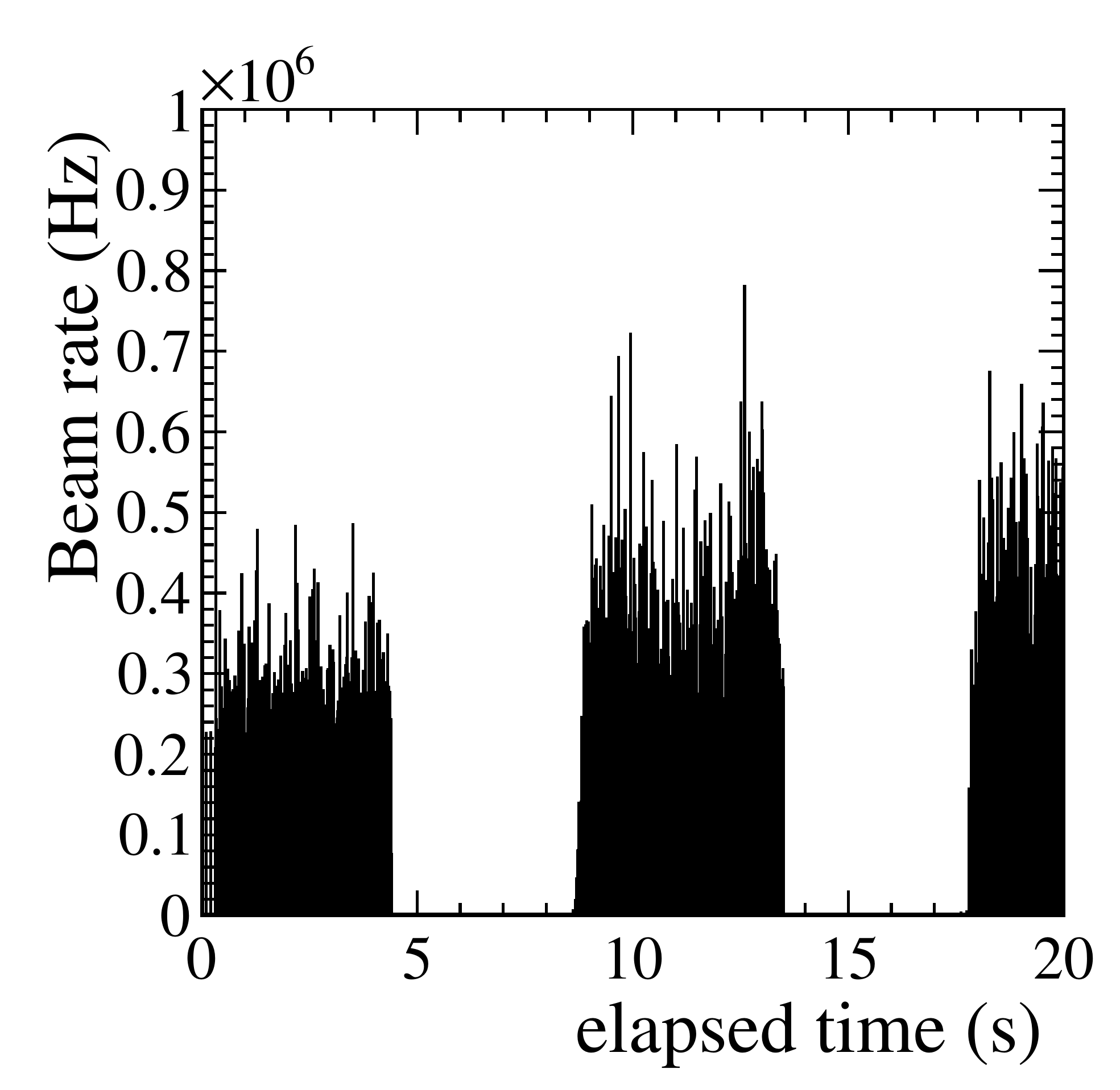}
\end{minipage}
\caption{The beam time profile for \he~ (left), \car~ (center) and
\oxi~ (right) data samples as a function
of the elapsed data taking time.}
\label{fig:spillstr}
\end{figure}

The beam time profile (\textit{Beam rate}) is shown for \he~(left), \car~(center) and
\oxi~(right) data samples in Fig.~\ref{fig:spillstr} as a function
of the elapsed data taking time (\textit{elapsed time}). The spill structure
of the beam is clearly visible.

\begin{figure}[h]
\centering
\begin{minipage}{0.5\linewidth}
\centering
\includegraphics[width = 1  \textwidth]{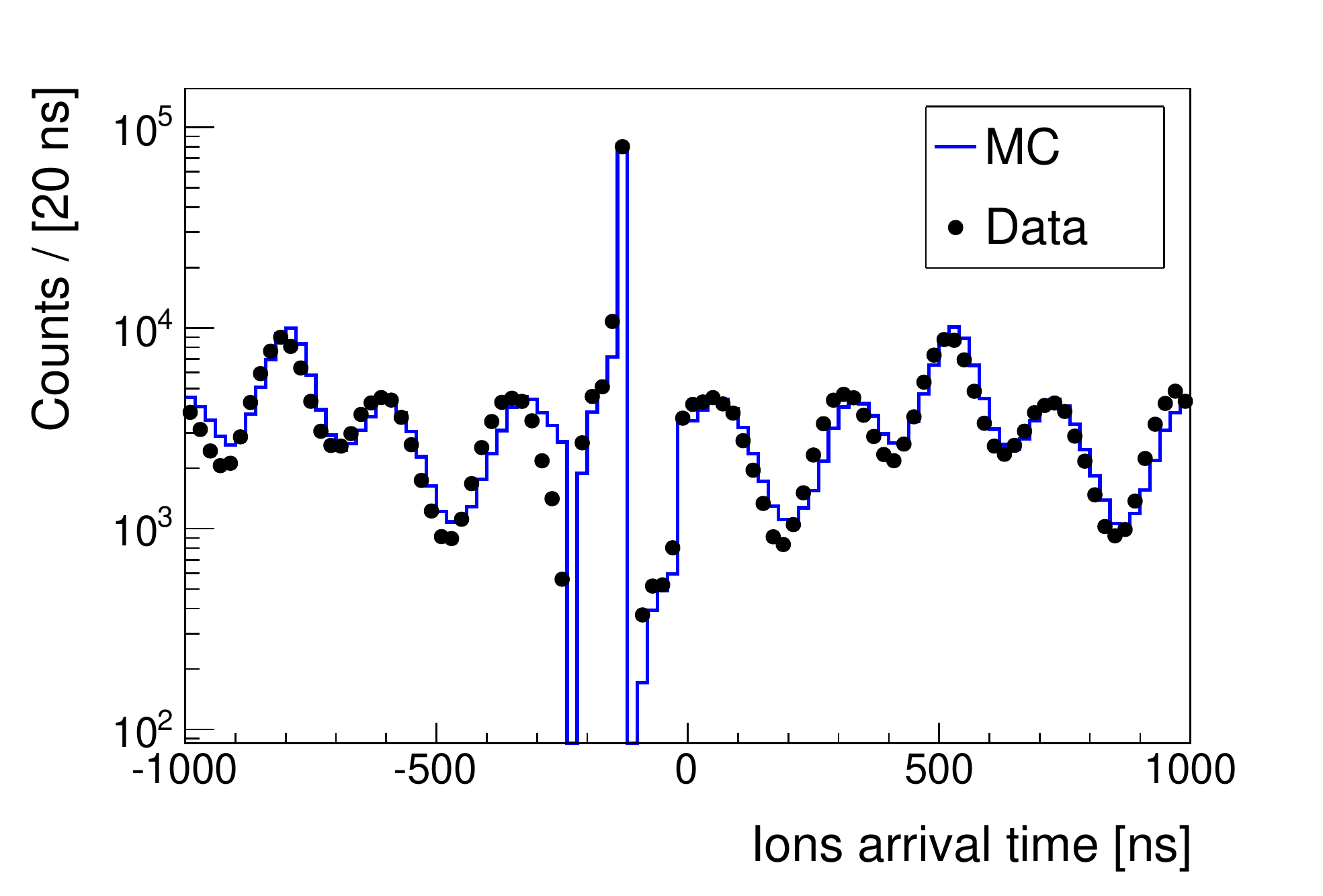}
\end{minipage}\hfill
\begin{minipage}{0.5\linewidth}
\centering
\includegraphics[width = 1  \textwidth]{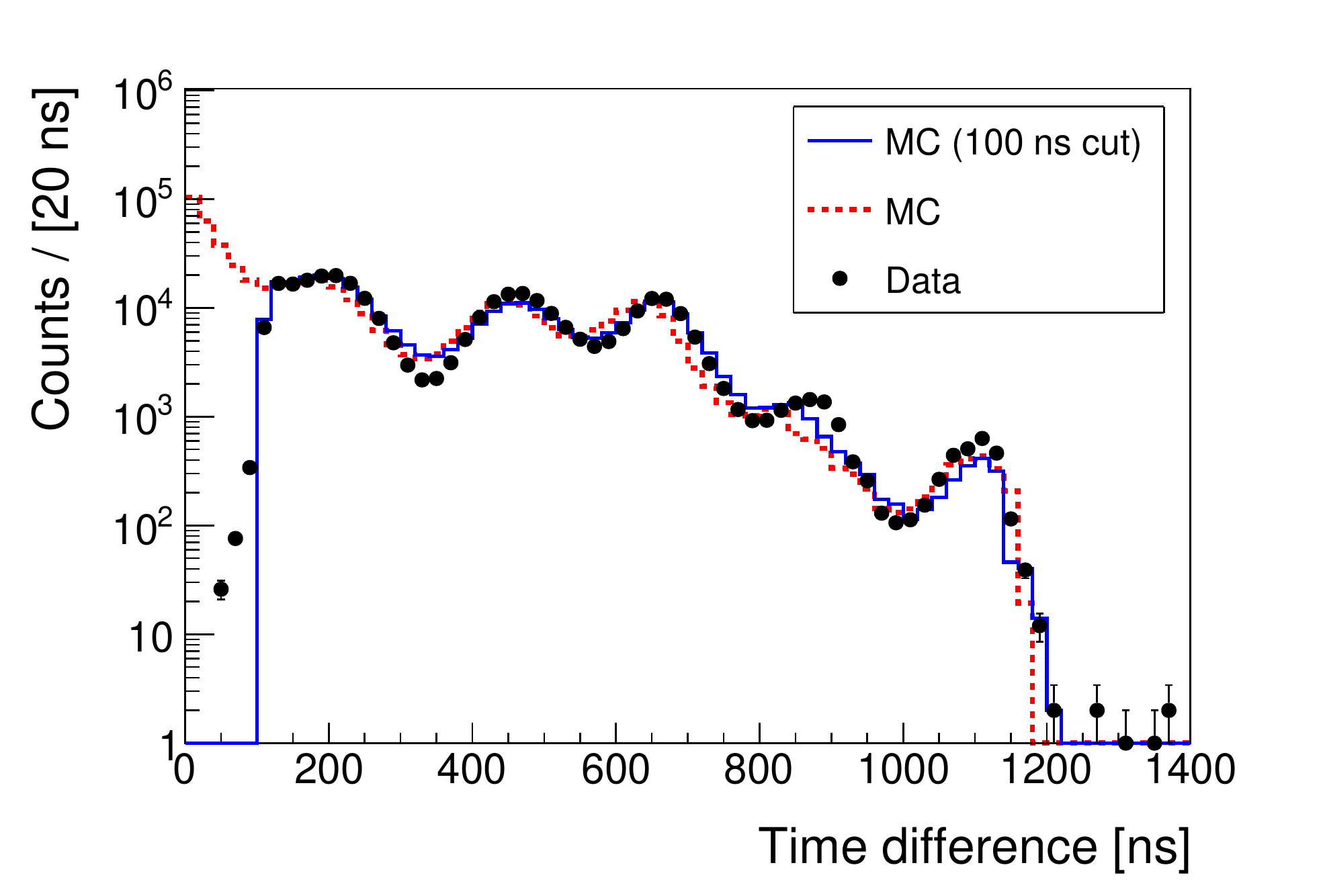}
\end{minipage}
\caption{Ions arrival time (left) and time difference between the detection of multiple ions (right)
are shown for events collected in a time window of $2\ \micro\second$ around the trigger time using the \he~beam of 
$144.63\ \mega\electronvolt$.
The data measured distributions are shown in black dots. The output of the dedicated MC simulation
used to account for discrimination induced inefficiencies, is superimposed in blue (solid line).
The output of a MC simulation that was not including the $100\ \nano\second$ discrimination time window is also shown,
in the left figure, in a red dashed line to show the impact on the multiple ions arrival detection.}
\label{fig:timespec}
\end{figure}

The fine structure of the beam, instead, has been measured by means of the multi hit TDC in 
a $2\ \micro\second$ window around the trigger time, and it is shown in black dots in Fig.~\ref{fig:timespec} (left)
for events collected using the \he~beams of $144.63\ \mega\electronvolt$. Similar spectra have been obtained 
also for the \car~ and \oxi~ beams. 
Using the beam fine structure measured in a time window far from the trigger, e.g. for the
\he~ events in the $200$--$800\ \nano\second$ window, a MC simulation has been performed
in order to account for the inefficiencies, in the detection of multiple ions events,
introduced by the fixed time window set by the 
discriminators used to process the analogic signals from the SC detector.
The ions rate used to generate the events in the MC simulation has been tuned
to optimize the agreement between the MC 
(shown as histograms in blue continuos lines)
and the data (shown as black dots) measured distributions shown
in Fig.~\ref{fig:timespec}. A remarkable agreement is observed,  allowing a 
reliable measurement of the discrimination window induced inefficiencies.

The difference in time arrival of multiple ions is shown in Fig.~\ref{fig:timespec} (right),
in black dots for the collected data, and shows the inefficiency introduced for
ions impinging on the SC with a time distance that is smaller than $100\ \nano\second$. 
The red dashed line in Fig.~\ref{fig:timespec} (right) shows the difference in 
arrival time of all the simulated ions, while the blue continuous line shows the
time difference after the $100\ \nano\second$ discrimination window is taken into account.
The systematic uncertainties related to the inefficiencies measurement are discussed
in \S~\ref{res}.

The beam spot size $B_{FWHM}$, as reported in Table~\ref{tab:datatake}, is inversely proportional to the beam energy for each ion species. It ranges between $6.9$--$9.3\ \milli\meter$ for \he~beams, $4.7$--$7.9\ \milli\meter$ for \car~beams and $3.6$--$4.6\ \milli\meter$ for \oxi~beams.

The maximum trigger rate was $6\ \kilo\hertz$, limit set by the DAQ dead time. 
The trigger lines were set up for the different measurements pointed out at the beginning of $\S$~\ref{exp-setup}: forward fragmentation studies, charged particles and prompt photons production at large angle and $\beta^+$ emitters production. 
The trigger line used for the prompt photon
studies had a mean rate in the $300\ \hertz$--$2\ \kilo\hertz$ range, depending on the beam
conditions, and contributed for the $5\%$--$30\%$ of the total trigger rate.


\section{Prompt photon yield}
\label{Yields}

The yield of prompt photons produced by the ion beam
projectiles interaction with the PMMA target, normalized to the total 
solid angle and integrated over the full target length, has been 
computed according to the following equation:\\
\begin{equation}
\label{eq:flux}
\Phi_{\rm \gamma} [\rm sr^{-1}] = \frac{1}{4\pi}\frac{N_{\rm \gamma}}{N_{\rm
    ion} \times \varepsilon_{\rm TOT} \times \varepsilon_{\rm DT} }
\end{equation}
\quad

where $N_{\rm \gamma}$ is the number of prompt photons
measured by the LYSO detector in the $(2-10)\ \mega\electronvolt$
energy range, 
$N_{\rm ion}$ is the total number of primary ions impinging on the 
PMMA target, $\varepsilon_{\rm TOT}$ is the total detector efficiency
(including the detector 
and geometrical contributions) and $\varepsilon_{\rm DT}$ is the
data acquisition dead time efficiency. 

Although the analysis has been performed starting from 
$1\ \mega\electronvolt$, the lower limit in the energy range 
for the yield evaluation was conservatively set to $2\ \mega\electronvolt$. 
Above this threshold, the background contribution, mainly due to 
intrinsic radioactivity of LYSO, electronic noise and neutrons, becomes negligible.  
The energy upper limit in the yield evaluation
was set to $10\ \mega\electronvolt$: above that energy we expected no 
significant de-excitation gamma 
lines from carbon and oxygen nuclei or their residues produced by 
nuclear reactions (\textit{National Nuclear Data Center - www.nndc.bnl.gov}). 

\noindent 
\begin{figure}[htbp!]
\centering
\begin{minipage}{0.5\linewidth}
\centering
\includegraphics[width = 1.  \textwidth]{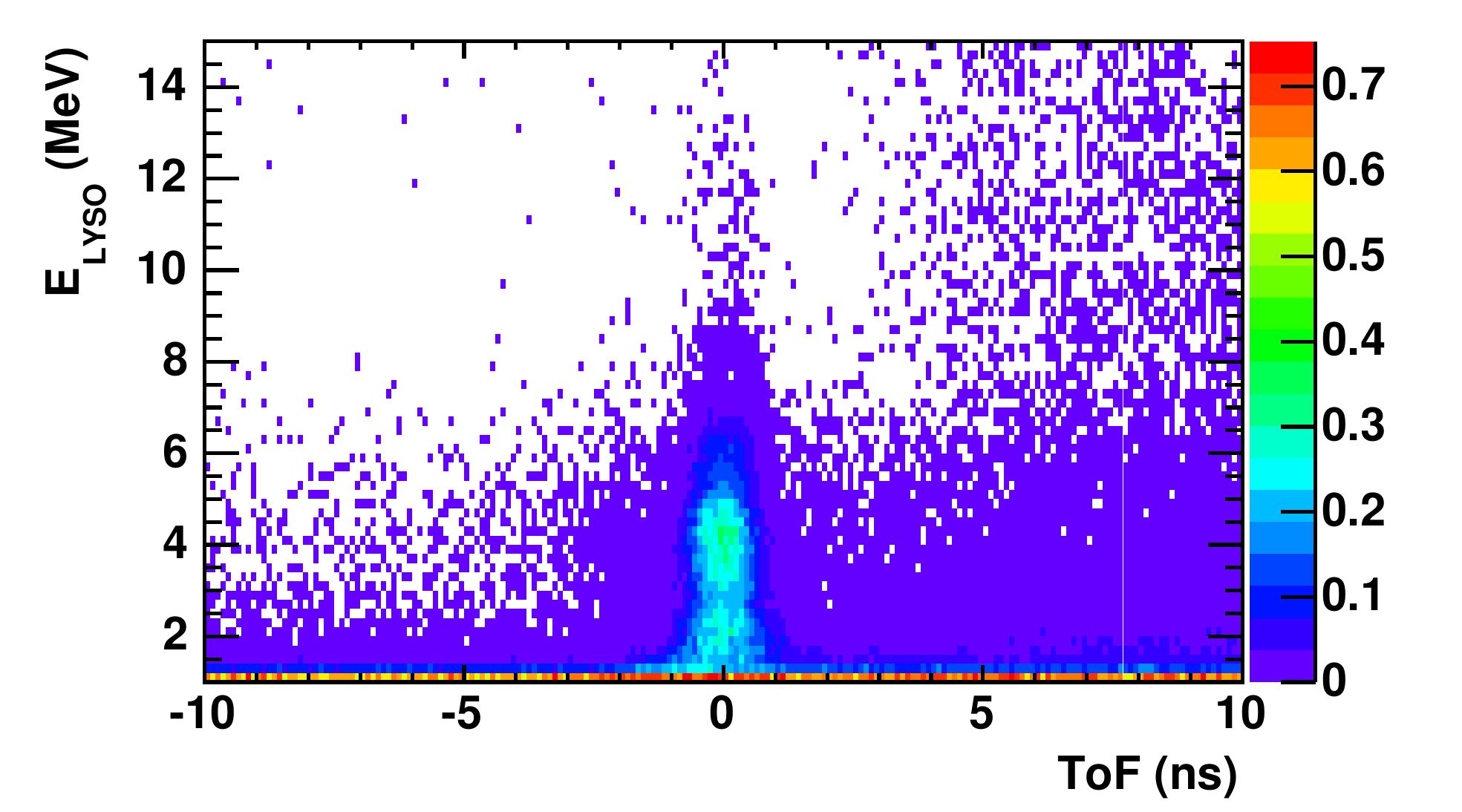}
\end{minipage}\hfill
\begin{minipage}{0.5\linewidth}
\centering
\includegraphics[width = 1.  \textwidth]{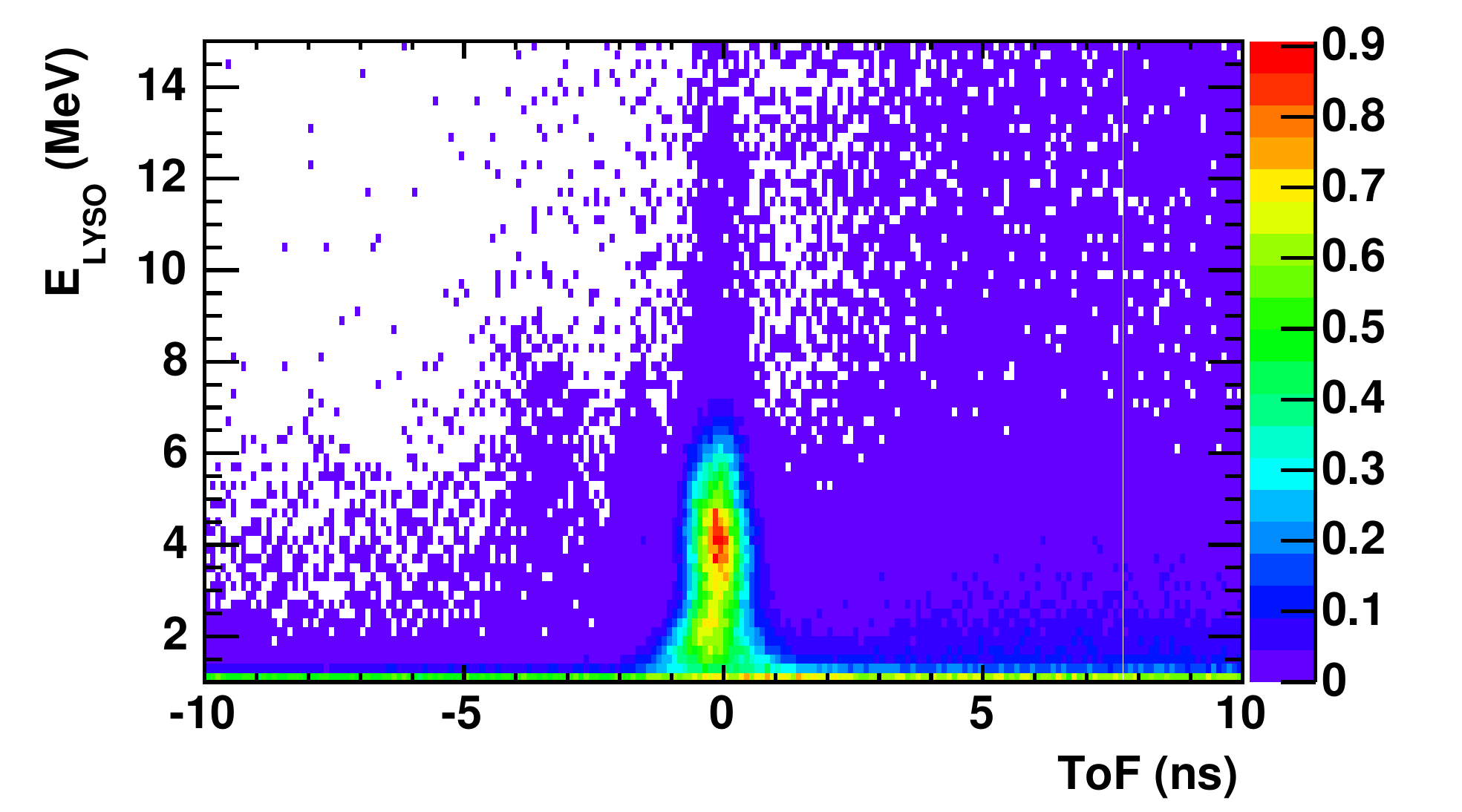}
\end{minipage}
\caption{Energy deposition in the LYSO crystal as a function of 
the measured ToF for \he~(left) and \oxi~ 
(right) events. The data collected at all energies, in the 90
degrees configuration, are summed up and shown for both beam types.
The distributions have been normalized to the same 
number (10$^7$) of incoming beam ions.}
\label{fig:elysovstof}
\end{figure}

The prompt photons number $N_{\rm \gamma}$ 
has been evaluated starting from the 2-dimensional distribution 
of the deposited energy in the crystal ($E_{\rm LYSO}$) as a function
of the Time of Flight (ToF) of secondaries interacting in the LYSO 
(see Fig.~\ref{fig:elysovstof}). 

The ToF is defined as the time 
difference between the signal detected in the LYSO and the 
signal detected in the SC (SCAND) induced by a traversing ion 
($T_{\rm LYSO} - T_{\rm SCAND}$). In this definition of ToF,
the time slewing effect is corrected following the same procedure described in~\cite{Agodi2012PP}.

Fig.~\ref{fig:elysovstof} shows, as an example, 
the $E_{\rm LYSO}$ vs ToF distribution used to select the prompt-$\gamma$ signal for the events taken using \he~ (left) and \oxi~ (right)
ions as projectiles. The distributions have been obtained
combining the data obtained in the different energy configurations
and collected at 90 degrees. A scaling factor has been applied
to both histograms in order to normalize the distributions to the
same statistics (10$^7$ events). 
The horizontal low energy band visible in both plots 
($E_{\rm LYSO} < 1.5$~MeV) is related to the 
LYSO intrinsic noise, while the diffused cloud is associated to 
neutrons, with a ToF that is almost uncorrelated to the 
SC signal. The vertical band at $0\ \nano\second$ is relative 
to the prompt photons signal.

\begin{figure}[htbp!]
\centering
\includegraphics[width = 0.9 \textwidth]{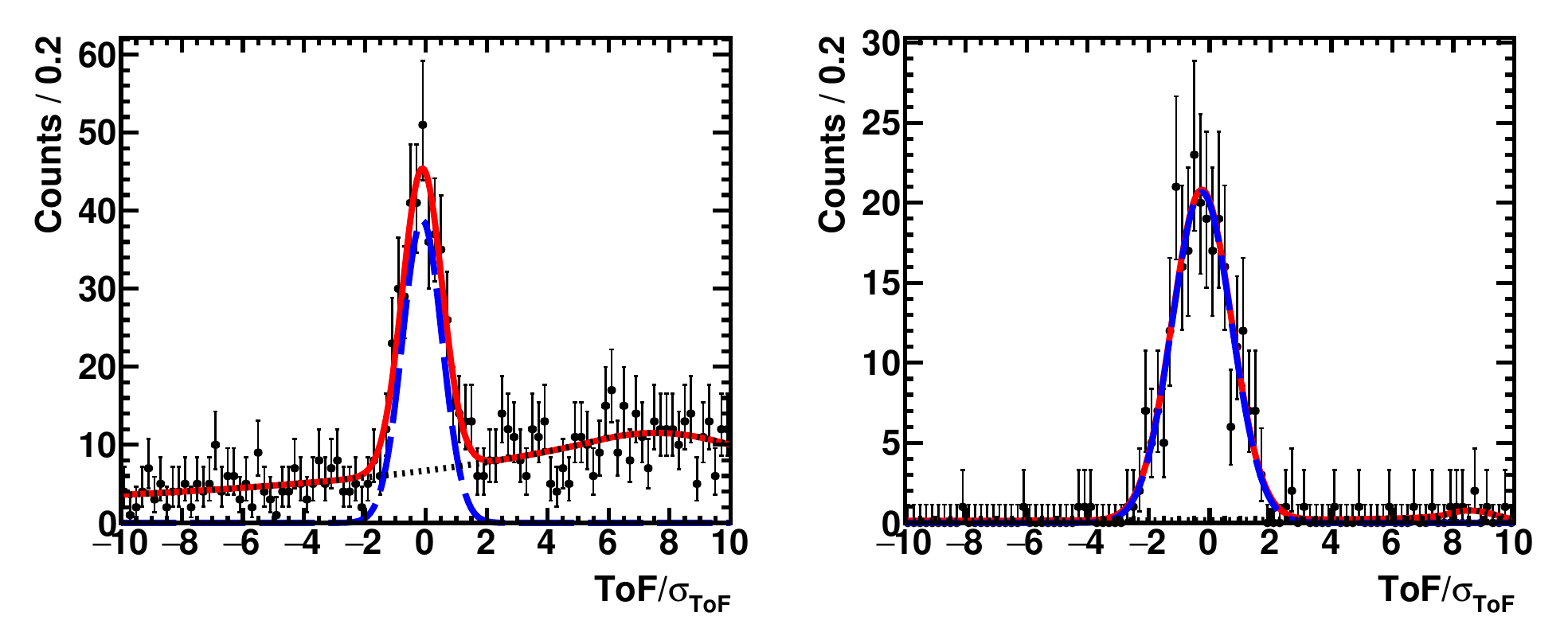}
\caption{Examples of the reduced time distribution in two different
  energy ranges (left:~$1.4\ \mega\electronvolt < \text{E}_{\rm LYSO}
  <1.6\ \mega\electronvolt$; right:~$4.6\ \mega\electronvolt<
  \text{E}_{\rm LYSO}<4.8\ \mega\electronvolt$) for a $145\
  \mega\electronvolt/\text{u}$ helium ion beam run in the $90\degree$
  angular configuration. 
The dashed line indicates the gaussian fit to the signal while the
dotted line is the Crystal Ball function fitting the neutron
background.}
\label{fig:Pullino_Lyso}
\end{figure}

$N_{\rm \gamma}$ has been computed from the reduced ToF 
(ToF/$\sigma_{\rm ToF}$) distributions, selected in the $[-10,10]$ 
interval and sampled in $E_{\rm LYSO}$ bins of $\sim 0.2\ 
\mega\electronvolt$. The $\sigma_{\rm ToF}$ values have been extracted from the time slewing correction procedure, where the ToF distributions sampled in $E_{\rm LYSO}$ bins have been modeled with a gaussian function (more details can be found
in~\cite{Agodi2012PP}). 
The number of prompt photons has been extracted from an 
unbinned maximum likelihood fit, performed using the RooFit toolkit
from ROOT~\cite{ROOFIT}, to the 
reduced time distribution for each energy bin. 
Fig.~\ref{fig:Pullino_Lyso} shows two examples of reduced time
distributions in lower ($1.4\ \mega\electronvolt < \text{E}_{\rm LYSO}
<1.6\ \mega\electronvolt$, shown in the left) and higher ($4.6\
\mega\electronvolt< \text{E}_{\rm LYSO}<4.8\ \mega\electronvolt$,
shown in the right) energy bins for a $145\
\mega\electronvolt/\text{u}$ helium ion beam run with the LYSO
detector placed at $\theta = 90\degree$.

The total fit function (solid line) is superimposed
to the data spectrum: the background,
mainly due to neutrons, is described by a Crystal Ball~\cite{Skwarnicki1986} shape (dotted
line), while the signal is modeled using a
gaussian Probability Density Function (PDF, dashed line). The Crystal Ball function well represents
the neutron background, especially in the low energy 
range ($E_{\rm LYSO} \sim 2\ \mega\electronvolt$). 
The background coming from low energy charged particles 
has been reduced by requiring that the number of 
hits detected in the drift chamber ($N_{\rm hit}$) is less than three
(the average number of hits detected for a charged particle traversing
the DCH is 12). As a cross-check of the charged particle background relevance
in the final result, the LTS detector has been used 
as an additional veto for charged particles: the combined use of
DCH and LTS did not result in a sensible change of the yields
and hence we did not explictly include the LTS charged particles
veto in the final analysis.

The dead time (DT) efficiency $\varepsilon_{\rm DT}$ has been evaluated 
using the VME scaler in which all the generated
trigger signals ($N_{TrTot}$) and the triggers signals acquired by the DAQ system ($N_{TrAcq}$), were counted.
The DT efficiency, defined as $\varepsilon_{\rm DT} =
N_{TrAcq}/N_{TrTot}$ varied from $60\%$ to $90\%$, depending on the beam
rate. Average values of $\varepsilon_{\rm DT}$ for the different data
taking conditions (ion species, beam energies and angular
configurations) have been used to compute the integrated 
yields using Eq.~\ref{eq:flux}. 

The total number of primary ions impinging on the PMMA target is
defined as $N_{\rm ion} = 
N_{\rm ion}^{\rm raw} \times corr_{\rm ion}$ where $N_{\rm ion}^{\rm raw}$ is the 
number of primary ions computed counting the number of SCAND signals over threshold as described in \S~\ref{exp-setup}. 

The obtained counts have to be corrected by $corr_{\rm ion}$,
a correction factor that takes into account the inefficiency
due to the $100\ \nano\second$ dead time introduced by the width of the SC discriminated signals.

As discussed in \S~\ref{beam}, and shown in Fig.~\ref{fig:timespec} (right),
this correction has been evaluated as the average fraction of lost 
ions using a dedicated MC simulation tuned in order to 
reproduce the temporal beam structure of each data
set that was acquired. $corr_{\rm ion}$  was measured as the ratio of 
all the events impinging on the SC, without applying any cut (as shown 
in Fig.~\ref{fig:timespec} right, red dashed histogram), to the number of
events in which the arrival time difference of multiple ions was greater than
$100\ \nano\second$ (as shown 
in Fig.~\ref{fig:timespec} right, blue solid histogram).
The measured $corr_{\rm ion}$ values, computed for each data set to
take into account the different beam rates, are in the range $[1.0,1.5]$.

\subsection{MC simulation}
\label{MCreso}

The detector and geometrical efficiencies ($ \varepsilon_{\rm TOT} $)  
have been computed using a MC simulation based on
the FLUKA code, that implemented all the experimental configuration
(beam characteristics and experimental setup as described 
in \S~\ref{exp-setup} by Table~\ref{tab:datatake} and shown in Fig.~\ref{fig:expset}). 

 The total efficiency $\varepsilon_{\rm TOT}$ has been defined as the ratio of
$N_{\gamma}^{\rm meas}$ to $N_{\gamma}^{\rm gen}$, where $N_{\gamma}^{\rm meas}$ is
the number of prompt photons measured by the LYSO detector, after having applied the same signal selections performed in the experimental data analysis, and $N_{\gamma}^{\rm gen}$ is 
the number of prompt photons produced with energy $> 2\ \mega\electronvolt$ by the ion beam interacting with
the PMMA target. The computed $\varepsilon_{\rm TOT}$ ranges from $6.6\cdot
10^{-5}$ up to $8.6\cdot 10^{-5}$, depending on the different experimental
setup conditions.

The computed efficiency depends only on the modeling of the prompt
photon transport from the production point inside the PMMA target up to
the LYSO detector. There is no dependence of the measured efficiency 
on the nuclear models used in FLUKA to generate the prompt photons 
emission spectrum as we are only taking into account the attenuation due to the interactions 
with matter of the emitted photons (related to the positioning and composition of the 
target and experimental setup), averaging
on their production position and energy. Care has been taken in order to ensure 
that a significant statistics was collected for all the relevant energies and 
emission positions allowing a measurement in which the
dominant contribution to the uncertainty was systematic.

As the interactions with matter 
of the photons in the 1--20 MeV energy range
is know to be very well reproduced in FLUKA, particular care has been made
in setting up a simulation in which all the detector and experimental apparatus
parts were properly included and positioned, especially those traversed by
the photons in their path towards the LYSO detector. The uncertainty related
to the geometrical survey measurements and detector description details
on the efficiency measurements are discussed in the systematic uncertainty
section (\S~\ref{res:sys}).

\subsection{Results}
\label{res}

\begin{figure}[htbp!]
\centering
\includegraphics[width = 0.9 \textwidth]{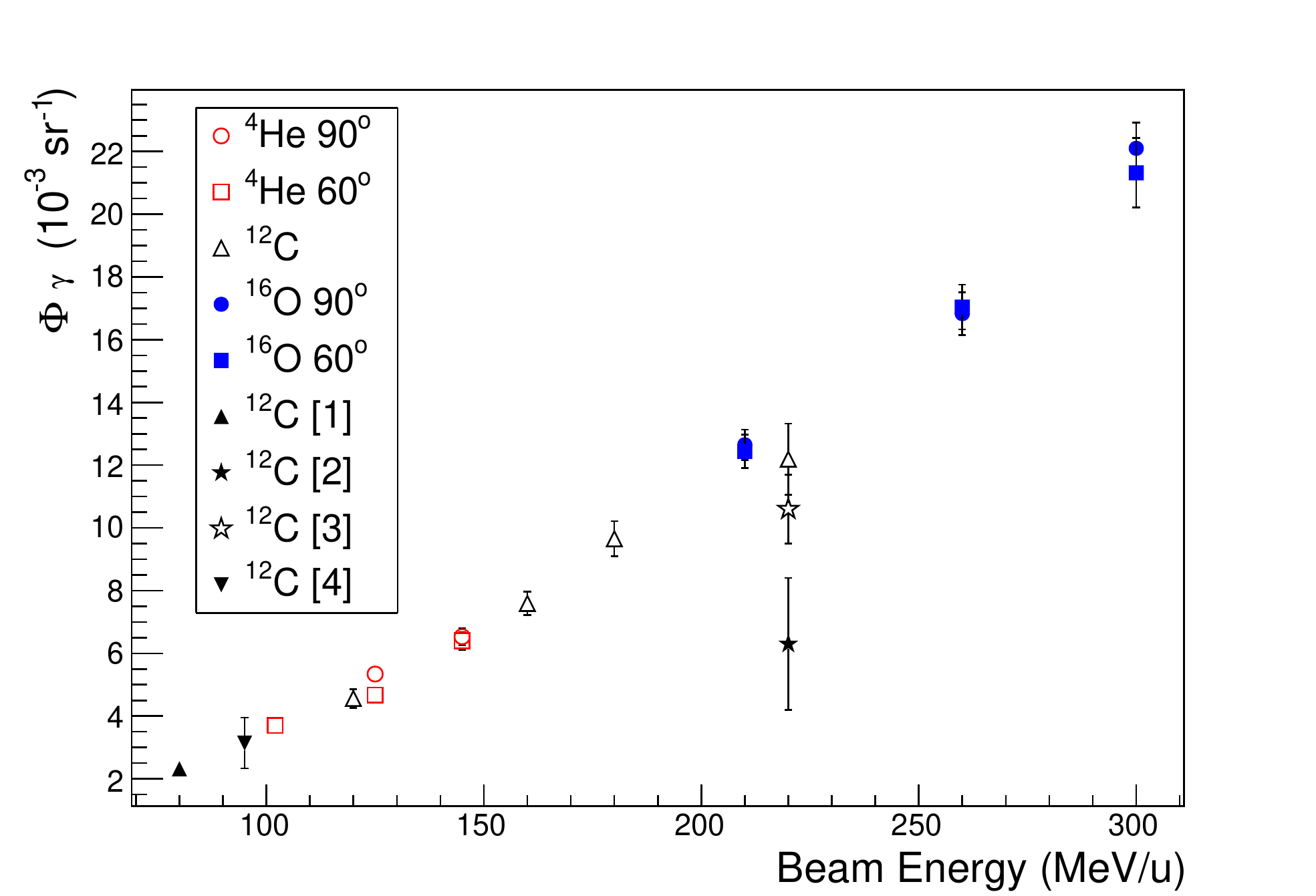}
\caption{Production yields of prompt photons as a function of the primary beam energy for helium (open circle - $90\degree$, open square - $60\degree$), carbon (open up triangle) and oxygen (full circle - $90\degree$, full square - $60\degree$). The results from references [1]~\cite{AgodiERRATA}, [2]~\cite{GsiLYSO220}, [3]~\cite{GsiBAF} and [4]~\cite{Pinto2015} are superimposed.}
\label{fig:fluxfig}
\end{figure}

The production yields of prompt photons ($\Phi_{\gamma}$) produced by \he~,\car~and \oxi~ion beams, computed according to
Eq.~\ref{eq:flux}, measured with the detector in the angular
configuration at 
$90\degree$ and $60\degree$, with a deposited energy E $>2\
\mega\electronvolt$, integrated over the full target size and
averaged in a full $4\pi$ solid angle, for the different ion beam 
energies are listed in Table~\ref{tab:yields}. For each ion species, $\Phi_{\gamma}$ increases with increasing energy, both at $90\degree$ and $60\degree$, while no angular dependence 
is evidenced in the integrated production yield. 
Furthermore, as a first approximation, an universal behavior as a function of the primary energy exists independently of the nuclear species, as shown in Fig.~\ref{fig:fluxfig}.

It has to be noted that the results presented in this manuscript
are related to the total integrated production of prompt photons occuring
as a direct consequence of the interactions of primary ions
with the target medium (primary component) that is limited
up to the BP region and of the other
indirect nuclear processes (secondary component) that can lead to 
a significant prompt photons production even after the BP region ~\cite{Pinto2015}.
While the data collected with the most energetic \car~ beam and the
\he~ and \oxi~ beams had a BP position that was close (1~cm or less)
to the exit (rear) PMMA face, and hence the secondary production 
of photons can be considered as similar and subdominant in all cases,
particular care has to be taken when comparing the low energy carbon
yields with results obtained in other experimental conditions.
However, the nice agreement of the yields shown in Fig.~\ref{fig:fluxfig}
for the \car~ beam at 120 MeV/u and the measurement presented 
in~\cite{Pinto2015} (\car~ at 95 MeV/u) and in~\cite{AgodiERRATA}  
(\car~ at 80 MeV/u) obtained in completely different experimental conditions
are suggesting that the secondary component,
while certainly present, is not dominant for 
those beam and (low) energy pair conditions.

\begin{table}[htbp]
\centering
{
\renewcommand\arraystretch{1.2}
\begin{tabular}{c c c c}
\centering
$\bm{\theta}$ & \textbf{Ion} &\textbf{Energy} & $\Phi_{\gamma}$ $\pm$ $\sigma_{\rm stat} $ $\pm$ $\sigma_{\rm sys} $\\  
  & &  (MeV/u) & $ (10^{-3}~sr^{-1})$
   \\  
\hline
\multirow{9}*{$90\degree$} & \multirow{2}* {$^{4}{\rm He}$} & 125 & $5.34 \pm  0.06 \pm 0.23 $\\ 
& & 145 & $6.53 \pm 0.07 \pm 0.27 $\\
\cline{2-2}\cline{3-3}\cline{4-4}

& \multirow{4}*{$^{12}{\rm C}$} & 120 & $4.56  \pm 0.09 \pm 0.28  $\\ 
&  & 160 & $7.59  \pm 0.13 \pm 0.35 $\\  
& & 180 & $9.65  \pm 0.18 \pm 0.53  $\\  
& & 220 & $12.19  \pm 0.24 \pm 1.11 $\\  
\cline{2-2}\cline{3-3}\cline{4-4}

& \multirow{3}*{$^{16}{\rm O}$} & 210 & $12.65 \pm 0.12\pm 0.47 $\\  
& & 260 & $16.83 \pm 0.20 \pm 0.65  $ \\  
& & 300 & $22.10 \pm 0.15 \pm 0.81 $ \\  

\hline

\multirow{6}*{$60\degree$} & \multirow{3}*{$^{4}{\rm He}$} & 102 & $3.70 \pm 0.08 \pm 0.11 $  \\  
& & 125 & $4.67 \pm 0.07 \pm 0.18 $  \\
& & 145 & $6.40 \pm 0.08 \pm 0.27 $  \\  
\cline{2-2}\cline{3-3}\cline{4-4}

& \multirow{3}*{$^{16}{\rm O}$} & 210 & $12.44 \pm 0.13 \pm 0.51  $ \\  
& & 260 & $ 17.04 \pm 0.19 \pm 0.69   $ \\ 
& & 300 & $21.32 \pm 0.19 \pm 1.09 $ \\
\hline
\end{tabular}
}
\caption{Production yields ($\Phi_{\gamma}$) of prompt photons,
  computed according to Eq.~\ref{eq:flux}, measured with the detector in the angular
configuration at  $90\degree$ and $60\degree$, with a deposited energy
threshold E $>2\ \mega\electronvolt$, integrated over the full target
size and averaged in a full $4\pi$ solid angle, for the different
ion beam energies. The statistical ($\sigma_{\rm stat}$) and systematic
($\sigma_{\rm sys}$) uncertainties are shown as well.} 
\label{tab:yields}
\end{table}

The results from carbon ion beams have been compared with other results from literature.
As reported in~\cite{GsiLYSO220}, the prompt photon production yield obtained from a $220\
\mega\electronvolt/\text{u}$ $^{12}$C beam impinging on a PMMA target,
integrated over the full target size ($5\times 5\times 20\ \centi\meter^{3}$) and
averaged in a full $4\pi$ solid angle, measured with a LYSO detector
at the GSI (Darmstadt, Germany) facility is
$\Phi_\gamma(^{12}\rm{C}~220~MeV/u~,E>2~MeV~@90\degree)  = (6.3 \pm
0.2_{\rm stat} \pm 2.1_{\rm sys}) \times 10^{-3}\ sr^{-1}$. 
This experimental result is less than 3 standard deviation away from the correspondent 
HIT yield $\Phi_\gamma$ shown in Table~\ref{tab:yields}.

The production yield was also measured at GSI using a BaF$_{2}$
detector~\cite{GsiBAF} obtaining a value of
$\Phi_\gamma(^{12}\rm{C}~220~MeV/u~,E>2~MeV~@90\degree)  = (10.6 \pm
0.1_{\rm stat} \pm 1.1_{\rm sys}) \times 10^{-3}\ sr^{-1}$, in agreement within 1 standard deviation with the 
corresponding value shown in Table~\ref{tab:yields}. 

The measured yields were also compared with what presented in 
\cite{Pinto2015}, where the prompt-$\gamma$ absolute yield produced by a
$95\ \mega\electronvolt/\text{u}$ $^{12}$C beam interacting with a
PMMA target, with an energy threshold of $2\ \mega\electronvolt$, averaged
over the full beam range, was reported to be $(1.74 \pm 0.09_{\rm
  stat} \pm 0.50_{\rm  sys}) \times 10^{-4}\
\text{mm}^{-1}~\text{sr}^{-1}$. 
In order to relate the integrated yield produced by the
$120\ \mega\electronvolt/\text{u}$ carbon ion beam shown in
Table~\ref{tab:yields} with the Pinto's result, we used an estimate of the full projected
range for our beam of $30.86\ \milli\meter$, as estimated by SRIM 2013.
With this assumption, the normalized rate becomes
$\Phi_\gamma(^{12}\rm{C}~120~MeV/u~@90\degree)  = 
(1.48 \pm 0.03_{\rm stat} \pm 0.09_{\rm sys}) \times 10^{-4}\ mm^{-1}~sr^{-1}$,
in agreement with what measured in \cite{Pinto2015}.

\subsection{Systematic uncertainties}
\label{res:sys}

Several sources are contributing to the total systematic
uncertainty ($\sigma_{\rm  sys}$) on the integrated absolute yield.

The $N_{\gamma}$ value was computed using an unbinned maximum
likelihood fit of the reduced ToF distributions. In order to evaluate the systematics related to the signal and 
background fit models, a different approach to count the signal events has been used. The background 
contribution in the signal region (ToF/$\sigma_{\rm ToF} \in [-3, 3]$) has been extrapolated from the
unbinned fit of the SideBands (SB) of the ToF/$\sigma_{\rm ToF}$ spectra 
(ToF/$\sigma_{\rm ToF} \in [-10, -3) \cup (3, 10]$). Therefore, the
signal events have been computed as the total number of events in the signal region after the subtraction of
the background  as obtained from the SB extrapolation.
The difference between the values obtained for $N_{\gamma}$ from the 
SB subtraction method and the full fit analysis, significant only 
for the cases of \car~beams, has been
used as an estimate of the systematic source from the $N_{\gamma}$ evaluation and added in
quadrature to $\sigma_{\rm sys}$ in the computation of the \car~$\Phi_{\gamma}$. The relative contribution of the 
$N_{\rm \gamma}$ systematics to the total uncertainty varies
from $\sim 2\%$ up to $\sim 8\%$.

The measured total number of primaries $N_{\rm ion}^{\rm raw}$ impinging on the PMMA target has been
computed using the number of SCAND signals (N$_{\rm SCAND}$). In order to evaluate the systematic uncertainty
due the N$_{\rm SCAND}$ evaluation, an independent measurement of $N_{\rm ion}^{\rm raw}$ is necessary. 
It has been provided by the
trigger line of the PET photons measurement, since the PET trigger is not related to the SC detector, 
though it is related to the true number of ions interacting in the target. 
The correlation between the number of PET triggers (N$_{\rm PET}$) and the 
SCAND signals has therefore been studied and founded to be within the 
statistical uncertainty of N$_{\rm PET}$. Since this is the intrinsic limit of the systematic error on 
$N_{\rm ion}^{\rm raw}$, the statistical fluctuation of N$_{\rm PET}$ has been assigned as systematic 
uncertainty to the estimated number of primaries. The relative contribution to
the total uncertainty from this source ranges between
$\sim 3 - 6\%$.

As already discussed beforehand, the $N_{\rm ion}^{\rm raw}$ measurement has been
corrected taking into account the inefficiency in the selection of multiple ions
impinging on the SC at times smaller than 100~ns using the
correction factor $corr_{\rm ion}$. The details on the MC simulation
that has been setup to account for this inefficiency are given in \S~\ref{beam}.
The systematic uncertainty due to this MC correction has been evaluated by 
performing a dedicated study in which the 
beam rate and the spill shape of the MC model have been varied within the
uncertainties. The beam rate was changed from the value obtained by the 
data/MC agreement tuning to the mean value measured on data while the
beam shape, obtained from a fit to the data sample, has been varied to take
into account the uncertainties on the fit parameters and fit range. 
The relative contribution to the total uncertainty from this source varies
from $\sim 1\%$ up to $\sim 5\%$.

The systematic uncertainty
on the total efficiency has been computed by using the FLUKA
MC simulation and varying the simulated setup geometry, in
order to account for the uncertainty on the relative positions of
the different detectors and of the PMMA target, and beam
size. The overall contribution to the total systematic uncertainty
was found to be negligible in all cases.


The systematic error due to the maximum energy cut ($10\ \mega\electronvolt$) 
in the $N_{\rm \gamma}$ integral computation has been checked 
against the MC simulation: the fraction of photons above that threshold ($f_{\rm lost}$) is of the order of 
$2\tcperthousand$ for the Helium and Carbon cases and for the Oxygen beam of 
$210\ \mega\electronvolt$ at $90\degree$. For the other Oxygen configurations $f_{\rm lost}$ is $\sim 1\%$
and has been added to $\sigma_{\rm sys}$.
	
Another possible systematic source to the
yield evaluation is the $2\ \mega\electronvolt$ lower limit of the energy range 
for the $N_{\rm \gamma}$ integral computation, due to possible energy shifts 
depending on the LYSO calibration curve. A systematic on the LYSO calibration
has therefore been computed varying the calibration parameters of $3\sigma$ around their 
mean values and taking into account their correlation coefficient: it is found to be negligible.

The systematic contribution coming from the background 
rejection of low energy charged particles using the $N_{\rm hit} < 3 $ selection 
has been studied. Exploiting the MC simulation that reproduces the DCH number 
of hit data distribution, the raw $N_{\rm \gamma}$ integral has been computed asking 
for $N_{\rm hit} < 2$ and $N_{\rm hit} < 4$ and no significant variation has been observed.

\section{Prompt-$\gamma$ monitoring applications}
\label{monitor}

The research and development of particle therapy online
monitoring devices relies on the experimental knowledge of 
the abundance and energy spectrum of secondary radiation
produced by the incoming beam interaction with the patient
body. As stated in the previous sections, the prompt photon radiation 
can be used to monitor the 
beam range inside the target during a PT treatment.
The resolution achievable on the measured beam range depends on the
production yield of prompt-$\gamma$'s, their emission energy and the technology that
is used to detect them. 

In order to have an estimate of the resolution achievable on the beam range
using the prompt photons produced by the interaction of \he~,\car~and
\oxi~ion beams with a PMMA target, we used the results published
in~\cite{Smeets2012}, where the performances of a slit camera
recently developed by the IBA are documented.  

The results presented in~\cite{Smeets2012} have been obtained studying
the prompt-$\gamma$'s produced by the interactions of a $160\
\mega\electronvolt$ proton beam impinging on a PMMA target.
Figure $32$ in~\cite{Smeets2012} shows the range estimation standard deviation as a function 
of the number of irradiated primary protons as measured by the IBA slit camera.\\
 
The use of a result obtained studying the interactions of a proton beam
to assess the monitoring performances reachable 
in PT treatments with \he~,\car~and \oxi~ion beams is possible,
once the number of primary ions that have to be shot in order to 
deliver a given dose to a selected target volume (depth and dimensions fixed) is known.
 
Using a dedicated MC simulation it is possible to compute the number of primary ions
in all the different configurations of ions (\he~,\car~and
\oxi) and energies of interest.  The produced
prompt photons in a given slice can hence be computed
using the results presented in this manuscript and the detector performances
(that are independent of the secondary radiation production source)
properly assessed. 

Two important implications of the proposed strategy have to be clarified: 
\begin{itemize}
\item the number of primary ions that has to be shot in a real treatment
to any target volume has a strong dependence on the details
of the total tumor region under treatment. Beside the obvious dependence
 on the energy of the beam, for each slice under treatment the number of
ions that have to be shot depends on the slice position within the total 
target volume, in order to account for the dose pile up and, also, of the
RBE weight applied to each voxel. The assumptions that have been used to obtain
the results presented in this manuscript are outlined below.
\item the number of prompt photons that are emitted during the treatment
has a dependence also on the material that is present after the BP: a secondary
emission can occur, not directly related to the nuclear interactions
of the primary beam particles with the target volume. The results obtained
in this manuscript are related to the total prompt photons production,
not having any experimental mean to disentangle the two different production
mechanisms. Hence, the flux used to compute the final number of 
photons has a direct dependence on the experimental conditions
used to obtain the results presented in table~\ref{tab:yields}, as discussed
in \S~\ref{res}.
\end{itemize}

The study documented in this manuscript refers to two distinct and
well defined treatment configurations, studied using a dedicated FLUKA MC simulation. 
The first one to be used as a reference (a) while the second one 
represents a real case scenario as described by~\cite{Kramer2000} (b): 
\begin{itemize}
\item[a)]{$1\ \gray$ dose homogeneously distributed in a slice of $3\ \milli\meter$ centered on the 
Bragg peak position at a depth in PMMA of $\sim 85\ \milli\meter$;}
\item[b)]{$1\ \gray$ dose delivered to a tumor of $120\ \centi\meter^3$, divided in $39$ slices.}
\end{itemize}
The conclusions, that are related to the number of prompt photons 
generated in the well defined scenarios are an indication of the 
expected number of produced photons in conditions that are,
under some assumptions, not far 
from typical treatment conditions but are not meant to be used
as representative of the final performances attainable on generic
PT treatments, as those will have to be assessed using dedicated
full MC simulations.

In both cases, the number of primary \he~,\car~and
\oxi~ions of a given energy needed to deliver the desired dose has been evaluated
and is, in the \car~ case, of the same order of magnitude of the number of ions used in real treatments performed at CNAO.
The physical dose (1 $\gray$) has been chosen as a reference considering that the RBE weighted dose is of the order of $\sim 2$~RBE-$\gray$
and that in a PT treatment at CNAO, using two laterally-opposed carbon-ion beams, a value of an RBE-weighted dose of $70.4$~RBE-$\gray$ divided in 16 fractions of $4.4$~RBE-$\gray$ each was prescribed to a skull-base chordoma target volume ($\sim 2$~RBE-$\gray$ for each beam)~\cite{TPSbatti}.
We thus used the 1 $\gray$ value as representative of the total
RBE weighted dose, averaged against the SOBP region, for a given PT 
with heavy ions. Such assumption, and the related conclusions, 
can be translated to different total doses to account for different
details of the treatment under study using a dedicated MC simulation.
However the conclusions on the feasibility of prompt photons monitoring
do not change, as they are related to the order of magnitude of the 
emitted photons.

The expected prompt photons absolute yields have been estimated 
using the results reported in
Table~\ref{tab:yields}, in the $90\degree$ angular configuration.\\

For the \textit{a} configuration, the number of primaries predicted by the MC simulation and the
relative amount of prompt-$\gamma$'s 
produced are reported in Table~\ref{tab:gray}. 
As previously outlined,~\cite{Smeets2012} quoted the standard deviation on the estimated beam range using the prompt photon radiation as a function of the number of $160\ \mega\electronvolt$ primary protons irradiating a PMMA target. Hence, using the measurement of the prompt photon yield emitted by a $160\ \mega\electronvolt$ proton beam reported in~\cite{Pinto2015}, we computed the number of primary protons needed to produce the prompt-$\gamma$ absolute yields listed in the last column of Table~\ref{tab:gray}.
For \he~beam we obtained $\sim 4 \times 10^8$ primary protons, for \car~and \oxi~beams we obtained $\sim 2 \times 10^8$ protons and, hence, an expected standard deviation on the beam range estimation
that is less than $2\ \milli\meter$.\\
\begin{table}[htbp!]
\renewcommand\arraystretch{1.3}
\centering
\begin{tabular}{*{1}{ >{\centering\arraybackslash}m{3.cm}}*{1}{ >{\centering\arraybackslash}m{2.5cm}}*{1}{ >{\centering\arraybackslash}m{4.4cm} }}
\hline
\textbf{Beam} & \textbf{Number of Primaries} & \textbf{Absolute Yield\ \ ($\times10^5$ counts/sr)} \\
\hline
$^{4}$He $125$ MeV/u & $1.9\times 10^{8}$ & $(10.15 \pm 0.11_{\rm stat} \pm 0.44_{\rm sys})$ \\
\hline
$^{12}$C $220$ MeV/u & $4.4\times 10^{7}$ & $(5.36 \pm 0.11_{\rm stat} \pm 0.49_{\rm sys})$ \\
\hline
$^{16}$O $260$ MeV/u & $2.4\times 10^{7}$ & $(4.04 \pm 0.05_{\rm stat} \pm 0.16_{\rm sys})$ \\
\hline
\end{tabular}
\caption{Number of primary
  particles needed to deliver a physical dose of $1\ \gray$ in a slice
  of $3\ \milli\meter$ centered on the BP position ($\sim 85\ \milli\meter$
  depth) in a PMMA target as predicted from a FLUKA MC simulation. The
  corresponding prompt photons absolute
  yield is reported in the last column, computed using the measurements
  shown in Table~\ref{tab:yields}. } 
\label{tab:gray}
\end{table}
\noindent These preliminary results can be used as a basis to discuss the prompt
photons applications to beam range monitoring in the scenario of a real
treatment plan, described in the \textit{b} configuration. In 
\cite{Kramer2000} a total
number of $\sim 7\times 10^8$ carbon ions is needed to deliver $1\ \gray$ of
absorbed dose to a tumor of $120\ \centi\meter^3$ divided in $39$
slices. 
As an exercise, assuming that each energy slice is
irradiated with $\sim 2 \times 10^7$ \car~ions of $220\
\mega\electronvolt/\text{u}$, we computed the expected absolute prompt photon
yield per slice that is $(2.44 \pm 0.05_{\rm stat} \pm 0.22_{\rm sys}) \times 10^{5}$ counts/sr. 
If $125\ \mega\electronvolt/\text{u}$ \he~and $260\
\mega\electronvolt/\text{u}$ \oxi~ion beams are considered, to deliver the same physical dose of $2\times 10^7$ \car~ions a number of $\sim 8.6\times 10^7$ \he~and $\sim 1.1 \times 10^7$ \oxi~ is needed, producing a $\gamma$ yield of $(4.59 \pm 0.05_{\rm stat} \pm 0.20_{\rm sys}) \times 10^{5}$ and
$(1.85 \pm 0.02_{\rm stat} \pm 0.07_{\rm sys}) \times 10^{5}$ counts/sr
respectively.  
The expected resolutions when using a slit camera have been computed for
the real case scenario as done before for the \textit{a} configuration: the number of equivalent $160\
\mega\electronvolt$ primary protons needed to produce the predicted
prompt-$\gamma$ yields from \he~,\car~and \oxi~ions has been estimated. The obtained relative expected standard
deviation on the beam range estimation has values smaller than $3\
\milli\meter$ for \car~and \oxi~and smaller than $2\
\milli\meter$ for \he~. 

It has to be noticed that physical doses larger than
$1\ \gray$ can be delivered in a typical treatment
fraction, and hence our estimate of the achievable
resolution on the beam range can be considered as conservative. 
Furthermore, we would like to
point out that in hypofractionated treatments the relevance of a
possible real--time monitoring will be particularly significant. Our results support 
the conclusion that
monitoring techniques exploiting prompt photons are feasible in
particle therapy, providing a resolution on the beam range 
matching the clinical requirements.\\

To evaluate the monitoring performances in real treatment cases, it is necessary to take into account all the patient/treatment specific characteristics, like the tumor volume, its location inside the body and the different tissues that have to be traversed by the beam.
A systematic study of the impact on the prompt gamma monitoring performances of the target inhomogeneities will be performed in the future, testing different geometrical configurations and beam energies.
Furthermore, a substantial contribution to predictions when considering a real treatment scenario can also be provided from the Monte Carlo simulations, progressively updated with experimental data measurements. 

\section*{Conclusions}

The prompt photon production of \he~, \car~and \oxi~beams
interacting with a beam stopping PMMA target has been studied at the
HIT Heidelberg facility with beam energies of interest for
PT applications. The production yield measurements performed in this study using
\car~ions beam are found to be in
agreement with results obtained from other experiments. 

The \he~and \oxi~beams, whose prompt-$\gamma$ production is herein measured
for the first time, are particularly relevant for future PT applications.
The obtained results confirm that a non
negligible prompt photons production occurs in the interactions of
\he~and \oxi~beams of therapeutical energy with a PMMA target.

The measured yields have been used to compute the expected resolution
on the beam range in a typical treatment scenario, assuming the
performances of the IBA slit camera documented in~\cite{Smeets2012}.
Resolutions below $2-3\ \milli\meter$ are obtained in all the different
scenarios supporting the feasibility of a prompt photons based beam
range monitoring approach for PT using \he~, \car~or \oxi~particle beams.

\section*{Acknowledgements}
We would like to thank sincerely Marco Magi (SBAI Department) for his valuable effort in the construction of several mechanical supports of the experimental setup. This work has been partly supported by the ``Museo storico della Fisica e Centro di studi e ricerche Enrico Fermi''. The access to the test beam at the Heidelberg Ion-beam Therapy center has been granted by the ULICE European program. We are indebted to Prof.~Dr.~Thomas Haberer and Dr.~Stephan Brons for having encouraged this measurement, made possible thanks to their support and to the help of the whole HIT staff.

\section*{References}

\bibliography{Main_neutri_bib}


\end{document}